\documentclass[11pt]{cernrep}
\usepackage{graphicx,epsfig}
\newcommand{\ttt}[1]{\texttt{#1}}
\newcommand{\tsc}[1]{\textsc{#1}}

\newcommand{\code}[1]
  {\hspace*{0.075\textwidth}\begin{minipage}{0.9\textwidth}{\small\ttt{#1}}
\end{minipage}\newline}
\begin{document}

\title{A Les Houches Interface for BSM Generators}

\author{J.~Alwall$^1$, E.~Boos$^2$, L.~Dudko$^2$, M.~Gigg$^3$, M.~Herquet$^4$,
 A.~Pukhov$^5$, P.~Richardson$^{3,6}$, A.~Sherstnev$^{7,2}$, 
P.~Skands$^{6,8}$} 
\institute{
$^1$SLAC, 2575 Sand Hill Road, Menlo Park, CA 94025-7090, USA\\
$^2$SINP, Moscow State University, Vorob'evy Gory, Moscow, 119992, Russia\\
$^3$IPPP, University of Durham, South Rd, Durham DH1
 3LE, UK\\
$^4$CP3, Universit\'e catholique de Louvain, 2 Chemin du Cyclotron,
  B-1348 Louvain-la-Neuve, Belgium\\ 
$^5$Faculty of Physics 1, Moscow State University, 
Leninskiye Gory, Moscow, 119992, Russia\\
$^6$CERN PH-TH, CH-1211 Geneva 23, Switzerland\\
$^7$Cavendish Laboratory, Cambridge University, 
Madingley Road, Cambridge CB3 0HE, UK\\
$^8$Theoretical Physics, Fermilab MS106, Box 500, Batavia, IL--60510,
USA
}
\maketitle

\begin{abstract}
We propose to combine and slightly extend two existing ``Les Houches
Accords'' to provide a simple generic interface between
beyond-the-standard-model parton-level and event-level generators. All
relevant information --- particle content, quantum numbers
of new states, masses, cross sections, parton-level
events, etc --- is collected in one single file, which adheres to the 
Les Houches Event File (LHEF) standard.
\end{abstract}

\section{INTRODUCTION}

The simulation of interactions at the LHC is characterized by the use of
many different programs specializing in different stages of
the calculation, such as matrix-element-level event generation, decay
of resonances, parton showering, hadronization, and underlying event
simulation. The communication of simulation parameters between those
stages can be complicated and program-specific. For supersymmetric
models, this situation has been greatly improved by the introduction
of the SUSY Les Houches Accord \cite{Skands:2003cj} (SLHA) and its upcoming
extension \cite{Allanach:2007zz}. For general models however, there is
still no corresponding standard. In this note, we suggest an addition
to the SLHA to allow for the specification of the
quantum numbers, masses, and decays of arbitrary new states, thus
generalizing the accord beyond its original supersymmetry-specific
scope. We also make a proposal for how to include these model
parameter files into Les Houches Accord event files 
\cite{Alwall:2006yp} (LHEF) in a standardized way. This 
both reduces the number of files that need to be passed
around and minimizes the possibility 
for error by keeping all relevant model information together with the
actual events. 

\section{DEFINITION OF THE INTERFACE}
The concrete proposal consists of the following three points:
\begin{enumerate}
\item Introduce new SLHA-like blocks \ttt{QNUMBERS} (for ``quantum
  numbers'') with the format: 
\begin{verbatim}
BLOCK QNUMBERS 7654321 # balleron
      1     0  # 3 times electric charge
      2     1  # number of spin states (2S+1)
      3     1  # colour rep (1: singlet, 3: triplet, 8: octet)
      4     0  # Particle/Antiparticle distinction (0=own anti)
\end{verbatim}
where this example pertains to a fictitious neutral spin-0
color-singlet self-conjugate particle to which we assign ``PDG'' code
7654321 and the name ``balleron''. That is, the \ttt{BLOCK}
declaration should define a PDG code and, optionally, a human readable
name after the \ttt{\#} character (if no name is given, the
PDG code may be used). 
We advise to choose PDG numbers in excess of 
3 million for new states, to minimize the  possibility of
conflict with already agreed-upon numbers \cite{Yao:2006px}. 
The entries so far defined are: \ttt{1}: the electric
charge times 3 (so that most particles will have integer values, but
real numbers should also be accepted); \ttt{2}: 
the particle's number of spin states: $2S+1$; \ttt{3}: the colour
representation of the particle, e.g., 1 for a 
  singlet, 3 (-3) for a triplet (antitriplet), 8 for an octet, etc.; 
\ttt{4}: particle/antiparticle distincition, should be 0 (zero) if the
particle is its own antiparticle, or 1 otherwise. 
\item Use the existing SLHA blocks
  \ttt{MASS} and \ttt{DECAY}  \cite{Skands:2003cj} 
  to define particle masses and decay tables. If the model in question
  is a SUSY model, a full SLHA spectrum 
  \cite{Skands:2003cj} can also be included. 
  We propose that the reader should ``turn on'' SUSY whenever
  the SLHA SUSY model definition block \ttt{MODSEL} is present. 
\item Include the information from points 1 and 2 enclosed within the subtags
  \ttt{<slha> </slha>} in the \ttt{<header>} part of Les Houches event files 
  \cite{Alwall:2006yp}. 
\end{enumerate}

\section{IMPLEMENTATIONS}
For the purpose of this contribution, the above proposal was tested
explicitly by interfacing \tsc{MadGraph/MadEvent} with
\tsc{Pythia}. Below we summarize the main aspects of these
implementations. 

\subsection{MadGraph/MadEvent implementation}
Starting from version 4 \cite{Alwall:2007st}, the multi-purpose
\tsc{MadGraph/MadEvent} parton-level event generator by
default includes a detailed summary of all simulation parameters in
the output LHEF \cite{Alwall:2006yp} parton-level event file. From
version 4.1.47, this information is stored in the XML
\texttt{<header>} section. For the interface
considered here, the relevant part of this section is a copy of the
so-called \ttt{param\_card.dat} \tsc{MG/ME} input file.

The \tsc{MG/ME} \ttt{param\_card.dat} uses an extension of the SUSY
Les Houches Accord \cite{Skands:2003cj,Allanach:2007zz} for model
parameters in all implemented models. In particular, it always
includes the \ttt{SMINPUTS}, \ttt{MASS}, and \ttt{DECAY} blocks. This
file is used by \tsc{MadGraph/MadEvent} as an input for cross
section computations and event generation but is not modified by the
program. The file is instead assumed to be created by an external
``Model Calculator''. Such calculators are currently available on the
web for the SM, MSSM and 2HDM models. Starting from the parameters in
the Lagrangian (primary parameters), they calculate all needed
secondary parameters (such as masses, decay widths, and auxiliary
parameters). Note that widths and branching ratios can also be
evaluated in an intermediate step by \tsc{MG/ME} itself or by 
external tools like \ttt{DECAY} or \ttt{BRIDGE} \cite{Meade:2007js}.

In previous versions of \tsc{MadGraph/MadEvent}, the
\ttt{param\_card.dat} file did not contain information regarding the
particle content of the physical model considered. This
information is stored in the \ttt{particle.dat} file filled by model
writers during the model creation. Starting from version 4.1.43, the
template for inclusion of user defined models (called \ttt{USRMOD}) in
\tsc{MadGraph/MadEvent} automatically generates the \ttt{QNUMBERS} blocks
described above from the information contained in the
\ttt{particle.dat} file. These blocks are then included in the default
\ttt{param\_card.dat} for the new model (and from there are copied
into the LHEF output), such that no extra
intervention is required to pass them to parton shower programs after
parton-level event production. The script only outputs information for
particles which have PDG numbers not identified as standard SM or MSSM
particles, since those are assumed to be defined in the parton shower
generators.

Note that in the current version, the spin, color and
particle/antiparticle information is automatically extracted, but not
the electric charge, which is set to zero by default. This is due to
the fact that, in \tsc{MadGraph/MadEvent}, the electric charge does
not appear in the list of particle properties and is only defined
through the value of the coupling to the photon. This issue will be
addressed in future versions of \ttt{USRMOD}, but can currently be
circumvented by fixing the electric charge information by hand at the
end of the model implementation process.

\subsection{Pythia implementation}
The following capabilities are implemented in \tsc{Pythia 6.414}
\cite{Sjostrand:2006za} and subsequent versions. 

Already for some time it has been possible 
to use the \ttt{QNUMBERS} blocks described above to define new
particles in \tsc{Pythia} via its SLHA interface \cite{Pukhov:2005je}.
What is new is that, when reading an LHEF event file,
\tsc{Pythia} now automatically searches for \ttt{QNUMBERS} blocks in the header
part of the LHEF file, updating its internal particle data tables
accordingly. It then proceeds to search for \ttt{MASS} and \ttt{DECAY}
tables, and finally looks for other SLHA blocks contained in the header. 
If the SUSY model definition block \ttt{MODSEL} is found, SUSY is
automatically switched on and the remaining SLHA blocks are read, without the
user having to intervene. The read-in of LHEF files containing general
BSM states, masses, and decay tables, should therefore now be relatively
`` plug-and-play''.  

A note on decay tables: only 2- and 3-body decays can currently be handled
consistently. They are then generated with flat phase space, according
to the branching ratios input via the \ttt{DECAY} tables. The colour
flow algorithms have been substantially generalized, but if too many
coloured particles are involved (e.g., an octet decaying to three
octets) \tsc{Pythia} will still not be able to guess which colour flow to
use, leading to errors. Please also read the warnings in the section
on decay tables in the SLHA report \cite{Skands:2003cj} 
concerning the dangers of
double counting partial widths and obliterating resonance shapes. 
To get around the
restriction to flat phase space, either 1) use \tsc{Pythia}'s 
internal resonance decays whenever possible 
(e.g., do not read in decay tables for particles for which
\tsc{Pythia}'s internal treatment is not desired modified), 2) perform
the decays externally, before the event is handed to 
\tsc{Pythia}, or 3) do a post facto re-weighting of the
generated events, based on the kinematics of the particle decays stored
in the event record. 

The interfaces can of course still also be used stand-alone,
independently of LHEF. The user must then manually open a 
spectrum file containing \ttt{QNUMBERS}
and \ttt{MASS} information and give \tsc{Pythia} the logical
unit number in \ttt{IMSS(21)}. 
New states can then be read in via either of the calls\\
\code{CALL PYSLHA(0,KF,IFAIL)~~~~! look for QNUMBERS for PDG = KF}
\code{CALL PYSLHA(0,0,IFAIL)~~~~~! read in all QNUMBERS}
and \ttt{MASS} information can be read by\\
\code{CALL PYSLHA(5,KF,IFAIL)~~~~! look for MASS entry for PDG = KF}
\code{CALL PYSLHA(5,0,IFAIL)~~~~~! read in all MASS entries}
where \ttt{IFAIL} is a standard return code, which is zero if
everything went fine. (For read-in of a complete SLHA SUSY spectrum file,
these direct calls should not be used, instead  set \ttt{IMSS(1)=11}
before the call to \tsc{PYINIT}.) For stand-alone decay table read-in, 
the unit number of the SLHA decay table file 
should be given in \ttt{IMSS(22)}, and the corresponding
read-in calls are\\
\code{CALL PYSLHA(2,KF,IFAIL)~~~~! look for DECAY table for PDG = KF}
\code{CALL PYSLHA(2,0,IFAIL)~~~~~! read in all DECAY tables}

\section*{CONCLUSIONS AND OUTLOOK}
We have proposed a simple file-based interface between parton- and
event-level generators focusing on the particular problems encountered
in the simulation of beyond-the-standard-model collider physics.  To
deal with general BSM models, we add a new block \ttt{QNUMBERS} to the
SLHA structure, which defines the SM quantum numbers of new states for
use in subsequent resonance decay, parton showering, and hadronization
programs. We also integrate the SLHA file into the existing LHEF
format to minimize the number of separate files needed.  The proposal
has been tested explicitly by implementations in the
\tsc{MadGraph/MadEvent} and \tsc{Pythia6} Monte Carlo event generators.

In the near future, also the \tsc{Herwig++} \cite{Bahr:2007ni} and
\tsc{Pythia8} \cite{Sjostrand:2007gs} generators will be extended to
automatically read in SLHA spectra from LHEF headers. Likewise,
forthcoming versions of the \tsc{CalcHEP} \cite{Pukhov:2004ca} and
\tsc{CompHEP} \cite{Boos:2004kh} parton-level generators will include
write-out of this information in their LHEF output, including also the
\ttt{QNUMBERS} extension.

In the longer term, with the XML format emerging as the de
facto standard for file-based interfaces, we note that it could be worth
investigating the merits of formulating an XML-SLHA scheme, that is,
transforming the current ASCII SLHA format conventions into a native XML
form that could be parsed with standard XML packages. A concrete 
first realization of such a strategy is HepML~\cite{Belov:2007qg} which
aims to unify the description of generator information in the form of
standard XML schemes, in which an XML-SLHA scheme would form a natural part.
The first release of the public HepML library 
has been implemented into CompHEP version 4.5, including 
also HepML headers in the LHEF output.

\section*{ACKNOWLEDGEMENTS}
The proposal and implementations contained herein 
originated at the workshop ``Physics at TeV Colliders'', Les Houches,
France, 2007.  This work has been partially supported by Fermi
Research Alliance, LLC, under Contract No.\ DE-AC02-07CH11359 with the
United States Department of Energy. The work of MH was
supported by the Institut Interuniversitaire des Sciences Nucl\'eaires
and by the Belgian Federal Office for Scientific, Technical and
Cultural Affairs through the Interuniversity Attraction Pole
P6/11. The HepML project is supported by the RFBR-07-07-00365 grant.
The work of JA was was supported by the Swedish Research
Council.

\bibliography{bsm-interface7}

\end{document}